\documentclass[preprint,12pt,longnamesfirst]{aastex}

\newcommand{\kms}{\hbox{ km\thinspace s$^{-1}$}}    %kms -1

\slugcomment{accepted for publication in AJ}

\shorttitle{Nebular Environment and X-ray Emission of KPD0005+5106}
\shortauthors{Chu et al.}

\begin{document}

%% LaTeX will automatically break titles if they run longer than
%% one line. However, you may use \\ to force a line break if
%% you desire.

\title{The Nebular Environment and Enigmatic Hard X-ray 
Emission of the Hot DO White Dwarf KPD\,0005+5106}

%% Use \author, \affil, and the \and command to format
%% author and affiliation information.
%% Note that \email has replaced the old \authoremail command
%% from AASTeX v4.0. You can use \email to mark an email address
%% anywhere in the paper, not just in the front matter.  
%% As in the title, you can use \\ to force line breaks.

\author{
You-Hua Chu\altaffilmark{1,2}, and
Robert A.\ Gruendl\altaffilmark{1},
Rosa M.\ Williams\altaffilmark{1,2},\\
Theodore R.\ Gull\altaffilmark{2,3},
Klaus Werner\altaffilmark{4}}
\altaffiltext{1}{Department of Astronomy, University of Illinois at 
Urbana-Champaign, 1002 West Green Street, Urbana, IL 61801}
\altaffiltext{2}{
Visiting Astronomer, Kitt Peak National Observatory,
National Optical Astronomy Observatories, operated by the
Association of Universities for Research in Astronomy, Inc. (AURA)  
under a cooperative agreement with the National Science Foundation.}
\altaffiltext{3}{Goddard Space Flight Center, NASA, Code 681,
Greenbelt, MD 20771-5302}
\altaffiltext{4}{Institut f\"ur Astronomie und Astrophysik, 
Universit\"at T\"ubingen, Sand 1, D-72076 T\"ubingen, Germany}
\email{chu@astro.uiuc.edu, gruendl@astro.uiuc.edu, 
rosanina@astro.uiuc.edu, gull@sea.gsfc.nasa.gov, 
werner@astro.uni-tuebingen.de}

%% Notice that each of these authors has alternate affiliations, which
%% are identified by the \altaffilmark after each name. Specify alternate
%% affiliation information with \altaffiltext, with one command per each
%% affiliation.

\begin{abstract}

We have detected an ionized nebula around the hot DO white 
dwarf KPD\,0005+5106, and used the [\ion{O}{3}]/H$\alpha$
ratios and nebular velocities to separate this nebula from
the background \ion{H}{2} region of AO Cas.
The angular size of the [\ion{O}{3}] nebula of
KPD\,0005+5106 is $\sim3^\circ$.
The velocity of this nebula is similar to those of the local 
interstellar \ion{H}{1} gas and the interstellar/circumstellar
absorption lines in UV spectra of KPD\,0005+5106, but has a 
large offset from those of the stellar photospheric lines.
The mass of the ionized nebula, $\sim70$ M$_\odot$, indicates
that it consists of interstellar material and that the velocity
offset between the star and the ambient medium should not
be interpreted as a wind outflow.
We have also analyzed the {\it ROSAT} PSPC observation of
KPD\,0005+5106 to determine its hard X-ray luminosity.
Using the $L_{\rm X}/L_{\rm bol}$ relation for late-type 
stars and the lack of obvious near-IR excess of KPD\,0005+5106,
we exclude the possible existence of a binary companion with 
coronal activity.
Without a wind outflow, the presence of stellar \ion{O}{8} 
emission requires that X-rays at energies greater than 
0.871 keV are present in the vicinity of KPD\,0005+5106.
This hard X-ray emission is most puzzling as neither 
photospheric emission at such high energies nor a 
high-temperature corona is expected from current
stellar atmospheric models KPD\,0005+5106.
X-ray observations with high angular resolution and sensitivity are
needed to confirm the positional coincidence and to enable
X-ray spectral analyses for constraining the physical origin of
the hard X-ray emission from KPD\,0005+5106.

\end{abstract} 

%% Keywords should appear after the \end{abstract} command. The 
%% uncommented example has been keyed in ApJ style. See the 
%% instructions to authors for the journal to which you are 
%% submitting your paper to determine what keyword punctuation 
%% is appropriate.


\section{Introduction}

About 20\% of white dwarfs (WDs) exhibit a deficiency or an absence
of hydrogen on their surfaces, and their evolutionary paths into 
such states are not well understood.
If planetary nebulae (PNe) are found around different types of WDs,
the nebular properties may shed light on the evolutionary history 
of the stars. 
Only hot WDs can photoionize their nebulae and make them shine in 
H$\alpha$.
Therefore, \citet{R87} searched for H$\alpha$ emission around 5 hot
WDs and 4 sdO stars using a Fabry-Perot spectrometer, but he found
emission only in the vicinity of the DO WDs PG\,0108+101 and 
PG\,0109+111, which are within $\sim1^\circ$ from each other.
Similarly, \citet{Wetal97} carried out a CCD imaging 
search for PNe around 16 hot WDs, but they detected H$\alpha$ 
nebulosity in the vicinity of only PG\,0109+111.
It was uncertain whether this ionized gas belonged to the 
interstellar medium or an evolved PN.
Among the non-detections, the hot DO star KPD\,0005+5106 presents
a puzzle because its far ultraviolet spectrum shows strong H$_2$
absorption lines indicating the presence of dense interstellar 
or circumstellar gas \citep{KW96,KW98}.
While this puzzle is partially solved by the recent detection of 
an \ion{O}{6} emission nebula by \citet{ODS04}, there remain many
other unusual properties of KPD\,0005+5106.

KPD\,0005+5106 has a stellar effective temperature of 
120,000 K \citep{WHF94}, which is hot enough to 
photoionize the \ion{O}{6}-emitting nebula.
It was the first WD reported to possess a corona heated by a wind,
as its X-ray spectrum below 0.5 keV could be satisfactorily fitted
only by emission models of He-rich plasma at temperatures of 
$2-3 \times 10^5$ K \citep{FWB93}.
This ongoing wind outflow has been used to explain the detection of
stellar \ion{O}{8} emission and circumstellar \ion{N}{5} absorption
in the high-resolution UV spectra of KPD\,0005+5106 taken with the 
{\it HST} GHRS \citep{Wetal96} and {\it IUE} SWP \citep{Setal97}.
These explanations should be reconsidered in light of the recent
discovery of hard X-ray emission from KPD\,0005+5106 near 1 keV 
\citep{Oetal03}.
This hard X-ray emission is able to photoionize O$^{+7}$ 
(ionization potential = 871 eV) to produce the observed \ion{O}{8}
recombination line emission, but the origin of this hard X-ray 
emission is not clear.
The hard X-ray spectrum of KPD\,0005+5106 indicates a plasma 
temperature greater than 10$^6$ K, which is too high to be
explained by the He-rich wind invoked by \citet{FWB93}.
Hard X-ray emission is frequently associated with WDs in binary 
systems and is attributed to the coronal activity of a late-type 
companion or the accretion of a companion's material onto the
surface of a WD; however, KPD\,0005+5106 does not show a near-IR 
excess and is not known to have a late-type companion \citep{Oetal03}.

To conduct a more sensitive search for a coronally active companion
of KPD\,0005+5106, we followed a similar approach to that used by 
\citet{Getal01} to illustrate the existence of a late-type dMe 
companion to the central star of the Helix Nebula.
KPD\,0005+5106 was monitored with high-dispersion spectra over 
four nights.
No variations in stellar features were detected, but moderately 
bright nebular emission was detected throughout the entire slit 
centered on KPD\,0005+5106.
The nebular emission around KPD\,0005+5106 consists of two velocity
components.  To determine the relationship between these nebular
components and KPD\,0005+5106, we have compared their velocities
to those of photospheric and interstellar absorption 
lines in UV spectra of KPD\,0005+5106, and examined images
from four emission-line surveys.  
We find that one nebular component is photoionized by KPD\,0005+5106
and the other component is associated with a background \ion{H}{2}
region.  We have also analyzed a {\it ROSAT} pointed observation of 
KPD\,0005+5106 to determine its hard X-ray luminosity and use it
to constrain the possible existence of a late-type companion.
In this paper, we describe the observations used in our investigation
of KPD\,0005+5106 (\S2); analyze the nebular environment (\S3), the
nature of the ionized nebula (\S4), and hard X-ray emission (\S5)
of KPD\,0005+5106; and discuss the implications of our results (\S6).

\section{Observations}

We have used four sets of wide-field emission-line images to
examine the morphological and spectral properties of the 
ionized medium around KPD\,0005+5106.
We have also used high-dispersion optical echelle observations of
KPD\,0005+5106 to search for variability of spectral features and 
to study the nebular emission.
Finally, archival {\it ROSAT} observations are used to quantitatively 
analyze the hard X-ray luminosity of KPD\,0005+5106.

\subsection{Narrow-Band Wide-Field Images}

Four sets of images of the field of KPD\,0005+5106 are available.
The first set of images are from an emission-line survey of the 
northern sky made by T.\ R.\ Gull and Y.-H.\ Chu in 1977-1979 using 
a Nikon 135-mm-focal-length lens and a Carnegie image tube with 
an RCA C33063 two-stage intensifier.  The camera was piggybacked 
on the No.\ 3 0.4-m telescope at Kitt Peak National 
Observatory (KPNO).
The instrumental setup was similar to that used by \citet{PGK79},
but the field-of-view was larger, $\sim$16$^\circ$, and the 
resulting plate scale was 26\farcm5 mm$^{-1}$.
KPD\,0005+5106, at $l$ = 116\rlap{$^\circ$}{.}1 and $b$ = 
$-$10\rlap{$^\circ$}{.}9, was well positioned in the survey field 
centered at $l$ = 115$^\circ$ and $b$ = $-$10$^\circ$.
Images were taken with H$\alpha$ and [\ion{O}{3}] filters for
720 s and 900 s, respectively, and recorded on hyper-sensitized Kodak
IIIa-J photographic plates.
The H$\alpha$ filter, centered at 6570 \AA\ with a FWHM of 75 \AA,
transmitted both the H$\alpha$ and [\ion{N}{2}] $\lambda\lambda$6548, 
6583 lines.
The [\ion{O}{3}] filter, centered at 5010 \AA\ with a FWHM of 28 \AA,
transmitted only the [\ion{O}{3}] $\lambda$5007 line.

The second set of images of KPD\,0005+5106 were taken by T.\ R.\ 
Gull in 1980 October as a follow-up study of interesting objects 
discovered from the Emission-Line Survey of the Milky Way \citep{PGK79}.
The instrumental setup differs from the one described above in that
a Nikon 300-mm-focal-length objective lens was used and the 
field-of-view was 7\rlap{$^\circ$}{.}1.
Images were taken with H$\alpha$ and [\ion{O}{3}] filters for
720 s and 1200 s, respectively.

A third wide-field H$\alpha$ image with higher 
sensitivity was available from the Virginia Tech Spectral-Line 
Survey\footnote{Available at http://www.phys.vt.edu/$\sim$halpha/.}.
The H$\alpha$ image from this survey used a 17 \AA\ wide
filter and thus rejected the flanking [\ion{N}{2}] lines.
The field of view was $\sim$10$^\circ$ in diameter, and the image
scale was 1\farcm6 pixel$^{-1}$.  
The detection limit of the H$\alpha$ image was 
$\gtrsim$ 2 pc cm$^{-6}$ in emission measure.

A fourth wide-field view of the region around KPD\,0005+5106 
was obtained from the Wisconsin H-Alpha Mapper Northern Sky
Survey (WHAM-NSS) data release version 1.0 \citep{H03}.
The WHAM-NSS data had a coarse spatial resolution, $\sim$1$^\circ$,
but recorded a high-resolution ($\sim$12 km s$^{-1}$) spectrum 
at each position.
We retrieved WHAM-NSS data for a large region centered at
KPD\,0005+5106.
The data were re-gridded spatially onto a common Galactic 
coordinate system with 10\arcmin\ pixels using a simple 
nearest-neighbor tiling algorithm.
The data were also interpolated and re-gridded spectrally onto 
a common local standard of rest velocity frame, i.e., 
$V_{\rm LSR}$, from $-$100 to +100 km~s$^{-1}$ with 
5 km~s$^{-1}$ pixels.
The final product is an image-velocity cube covering
$28^\circ\times28^\circ\times200$ km~s$^{-1}$.

\subsection{High-Dispersion Echelle Spectroscopy}

We obtained high-dispersion spectroscopic observations of 
KPD\,0005+5106 with the echelle spectrograph on the 4m telescope
at KPNO on 2003 November 4--7.  These observations were made with a 
79 line mm$^{-1}$ echelle grating, a 226 line mm$^{-1}$ 
cross-disperser and a broad-band blocking filter (GG385).  
The long-focus red camera was used to obtain a reciprocal 
dispersion of 3.5 \AA\ mm$^{-1}$ at H$\alpha$.  
The spectra were imaged using the T2KB CCD, of which the
24 $\mu$m pixel size corresponded to 0\farcs 26 pixel$^{-1}$ along 
the slit and $\sim$0.08 \AA\ pixel$^{-1}$ along the dispersion axis.  
A 2$''$ slitwidth was used, resulting in an instrumental 
resolution of 0.3 \AA\ (14~\kms\ at H$\alpha$), as measured by the 
FWHM of the unresolved telluric emission lines.
A slit length of 15$''$ was used for these multi-order observations
so that a wavelength coverage of 4500--7500 \AA\ was achieved.
The slit was north-south oriented and centered on KPD\,0005+5106.

Each night the echelle observation of KPD\,0005+5106 was split into
two 900 s exposures to aid in cosmic-ray removal.  
The observations were reduced using standard packages in 
IRAF\footnote{IRAF is distributed by the National
Optical Astronomy Observatories, which are operated by the Association
of Universities for Research in Astronomy, Inc., under cooperative 
agreement with the National Science Foundation.}.
All spectra were bias and dark corrected, and cosmic ray events 
were rejected.  Observations of a Th-Ar lamp were used for 
wavelength calibration, and the telluric lines in the source 
observations were used for a precise velocity alignment.
The accuracy of velocity measurements is $\pm$2 km~s$^{-1}$
in the H$\alpha$ + [\ion{N}{2}] order and $\pm$5 km~s$^{-1}$
in the [\ion{O}{3}] order.
Absolute flux calibration was not possible due to the full moon 
and variable clouds in the sky during the echelle observing run.
Fortunately, absolute flux calibration is not needed for 
extracting velocity profiles from the echelle data.

Nebular H$\alpha$, [\ion{O}{3}] $\lambda$5007, and [\ion{N}{2}]
$\lambda$6584 lines are detected at levels of 2--4 counts 
pixel$^{-1}$ above a sky background of 4--5 counts pixel$^{-1}$.
The velocity profiles of these three lines are extracted from
regions 5$''$ above and 5$''$ below the spectrum of 
KPD\,0005+5106 with the sky emission approximated as a constant
and subtracted.
No other nebular lines were detected in the echelle observations.

%{{\it FUSE} Observations}
%KPD\,0005+5106 has been observed at multiple epochs by the {\it FUSE}
%satellite as part of program M107 to monitor the wavelength
%calibration of the spectrographs.  The {\it FUSE} observatory has four 
%spectrographs operating simultaneously to cover the 905--1187 \AA\ 
%wavelength range with a resolution of $\lambda/\delta\lambda 
%\simeq $15,000--20,000.  Further details of the design and performance
%of the {\it FUSE} spectrographs are described by \citet{moos00} and 
%\citet{sahnow00}.

%We have retrieved {\it FUSE} observations of KPD\,0005+5106 from
%13 epochs taken in the time-tag mode.  
%Since we are primarily interested in the wavelength region
%around \ion{O}{6} we have concentrated our analysis on those obtained
%with the LiF~1a detector segment.
%These observations were reprocessed using the {\it FUSE} 
%calibration pipeline software CALFUSE (v2.4.0) to extract a 
%1-dimensional spectrum for each exposure.  A detailed description 
%of the pipeline processing is available in the {\it FUSE} Data 
%Handbook\footnote{Available at 
%http://fuse.pha.jhu.edu/analysis/dhbook.html.}.
 
\subsection{{\it ROSAT} X-ray Observations}

KPD\,0005+5106 was observed by the {\it ROSAT} Position Sensitive
Proportional Counter (PSPC) with a boron filter on 1991 December 31
for 5 ks (Observation number RF200428; PI: M.\ Barstow).  
The PSPC is sensitive to photon energies between 0.1 and 2.4 keV.
The boron filter, having little transmission between 0.188 and 
0.28 keV, was designed to provide additional spectral information
at this soft energy band, but it also reduced the transmission
at 1 keV by $\sim$ 20\%.

\section{Nebular Environment of KPD\,0005+5106}

We first noticed an ionized nebula around KPD\,0005+5106
from our echelle observations, which revealed nebular
line emission throughout the 15$''$ slit length.
Only the nebular H$\alpha$, [\ion{N}{2}] $\lambda$6583, 
and [\ion{O}{3}] $\lambda$5007 lines are well detected;
their extracted velocity profiles are shown in Figure 1.
The H$\alpha$ and [\ion{N}{2}] line 
profiles are similar, barring their different thermal 
widths and the contaminating geocoronal H$\alpha$ emission
near 0 km~s$^{-1}$ observed velocity, or  
$V_{\rm LSR}\sim3$ km~s$^{-1}$; both lines
show velocity components at $V_{\rm LSR}\sim-$7 and $-$32 
km~s$^{-1}$.
The [\ion{O}{3}] line, on the other hand, shows only one 
velocity component near $-$3 km~s$^{-1}$, which is most 
likely associated with the $-7$ km~s$^{-1}$ component of
the H$\alpha$ and [\ion{N}{2}] lines.
The drastically different [\ion{O}{3}]/H$\alpha$ ratios
between the $-7$ and $-$32 km~s$^{-1}$ velocity components 
indicate that these two components are associated with 
ionizing stars of very different effective temperatures.

The two sets of H$\alpha$ and [\ion{O}{3}] images taken 
with an image tube camera and a 135-mm or 300-mm objective
lens are shown in Figures 2 and 3.
Despite the different degrees of vignetting in the 
field-of-view, both sets of images show that the 
H$\alpha$-emitting region is more extended than the 
[\ion{O}{3}]-emitting region and that the [\ion{O}{3}] 
region is roughly centered on KPD\,0005+5106.
A deeper H$\alpha$ image is available from the Virginia
Tech Emission-Line Survey\footnote{Available at
http://www.phys.vt.edu/$\sim$halpha/fields/Cas01/.},
and it shows that 
KPD\,0005+5106 is superposed on the western part of an 
H$\alpha$-emission region about 4$^\circ$ across.

The iso-velocity maps made with the WHAM-NSS data provide
a complete view of the emission region at a lower
spatial resolution.
As shown in Figure 4, two spatially distinct emission 
regions near KPD\,0005+5106 can be identified with 
central velocities similar to the two velocity components
revealed in the H$\alpha$ and [\ion{N}{2}] line profiles
from our echelle observations.
The emission region associated with the $-$32 km~s$^{-1}$
component is brighter.
It is $\sim$4$^\circ$ across and best seen in the 
iso-velocity maps near $-$30 km~s$^{-1}$; its brightest 
patch of emission can be traced from $-$5 to 
$-$55 km~s$^{-1}$.
This $-$32 km~s$^{-1}$ emission region has been identified
to be the \ion{H}{2} region of AO Cas at a distance of
3.7 kpc \citep{R88}.
AO Cas is a binary system with the primary being an O9\,III 
star and the secondary being of a similar spectral type 
\citep{W73,GW91}.
The absence of [\ion{O}{3}] emission from this \ion{H}{2}
region (see Figs.~2b \& 3b) is consistent with the late-O 
spectral types of the AO Cas binary system.
Therefore, this \ion{H}{2} region is a background object
to KPD\,0005+5106.

The emission region associated with the $-$7 km~s$^{-1}$ 
component is faint, and the superposition of the emission 
from the \ion{H}{2} region of AO Cas makes it difficult to 
identify the boundary of this $-$7 km~s$^{-1}$ component.
The spatial extent of the $-$7 km~s$^{-1}$ emission region
is better seen in the iso-velocity maps at $V_{\rm LSR} 
\ge -10$ km~s$^{-1}$, where the contamination from the 
red wing of the $-$32 km~s$^{-1}$ component is negligible.
The brightest region of the $-$7 km~s$^{-1}$ component is
a 3$^\circ$-long horizontal band of emission below 
KPD\,0005+5106.
On the two ends of this band of emission, fainter filaments of
emission curve toward lower galactic latitudes and form an 
apparent shell structure about 10$^\circ$ across.
This $-7$ km s$^{-1}$ H$\alpha$ emission region is much more
extended than the [\ion{O}{3}] region shown in Figure 2b; only 
the brightest part of the H$\alpha$ emission region overlaps 
with the [\ion{O}{3}] region.
It is not clear whether the apparent shell structure is associated
with KPD\,0005+5106 or a result by chance superposition.
As KPD\,0005+5106 has an effective temperature of 120,000 K
\citep{WHF94}, its photoionized region should have high 
excitation with high [\ion{O}{3}]/H$\alpha$ ratios.
Therefore, we suggest that only the [\ion{O}{3}]-emitting 
region of the $-7$ km~s$^{-1}$ component is associated with 
KPD\,0005+5106.

\section{The Nature of the Nebula Photoionized by KPD\,0005+5106}

Is the $-$7 km~s$^{-1}$ ionized gas associated with 
KPD\,0005+5106 interstellar or circumstellar in origin?
The UV spectra of KPD\,0005+5106 show photospheric absorption 
lines at an average $V_{\rm LSR}$ of +42 km s$^{-1}$ and
interstellar/circumstellar absorption lines at an average 
$V_{\rm LSR}$ of $-$7 km s$^{-1}$ \citep[converted from the 
$V_{\rm hel}$ reported by][]{Wetal96}.
It might be hypothesized that the interstellar/circumstellar 
absorption arises from the approaching side of an expanding PN
of KPD\,0005+5106.
The systemic velocity of the PN should then be similar to the
velocity of the central star, unless the PN has been braked by
dense interstellar medium.
Adopting the average photospheric absorption line velocity of
KPD\,0005+5106 as the systemic velocity of the PN, the
expansion velocity of the PN would be nearly 50 km s$^{-1}$ 
and we would expect to see H$\alpha$ and [\ion{O}{3}] emission 
from the receding side of the PN at +90 km s$^{-1}$.
Our echelle observations do not show any nebular emission at 
this velocity; thus, we may rule out the expanding PN hypothesis.

It is likely that the ionized nebula of KPD\,0005+5106 consists 
of interstellar material.
The interstellar \ion{H}{1} in this direction spans a wide 
range of velocities, but the main component is at roughly
$-15 \le V_{\rm LSR} \le +5$ km s$^{-1}$ and has a smooth
spatial distribution on scales of 5$^\circ$--10$^\circ$
\citep{HB97}.
The smooth morphology and velocity distribution of the
interstellar \ion{H}{1} gas are both similar to those of the 
nebula photoionized by KPD\,0005+5106, supporting the 
interstellar origin of the nebula.

The size and mass of the nebula ionized by KPD\,0005+5106 
provide further information about its origin.
The distance to KPD\,0005+5106 has been determined 
spectroscopically to be 270$^{+210}_{-120}$ pc \citep{WHF94}.
The 3$^\circ$ angular size of the [\ion{O}{3}]-emitting region
corresponds to a linear size of 14$^{+11}_{-6}$ pc.
This size rivals that of ``the largest PN'' recently discovered 
around PG\,1034+001 by \citet{Hetal03} and its enormous ionized 
halo subsequently reported by \citet{Retal04}.
However, we find an interstellar origin to be more probable
for the nebula of PG\,1034+001 because its nebular velocity
\citep[$\sim$0 km s$^{-1}$,][]{Hetal03} is similar to those of 
the low-velocity local interstellar \ion{H}{1} gas \citep{HB97} 
and the circumstellar/interstellar absorption lines 
\citep[+7 km s$^{-1}$,][]{Hetal98}, but very different from 
those of the photospheric lines of PG\,1034+001 
\citep[+51 km s$^{-1}$,][]{Hetal98}.
Confirmed PNe are all much smaller than the ionized nebula
of KPD\,0005+5106.

To determine the mass of the ionized nebula of KPD\,0005+5106,
we use the WHAM-NSS H$\alpha$ line integrated over a 
2$^\circ$-radius aperture centered on KPD\,0005+5106.
The integrated H$\alpha$ line profile is fitted and decomposed
into two Gaussian components at $-$7 and $-$32 km~s$^{-1}$.
The H$\alpha$ flux of the $-$7 km s$^{-1}$ component is used
in the mass estimate.
Adopting a distance of 270 pc, we find that the total H$\alpha$
luminosity of the gas in the $-$7 km~s$^{-1}$ component to be 
$\sim$2$\times$10$^{34}$ ergs~s$^{-1}$.
Assuming a homogeneous Str\"omgren sphere, we find an rms density
of $\sim$0.8 H-atom cm$^{-3}$ and the total mass of the ionized 
nebula of KPD\,0005+5106 to be 70 M$_\odot$.
This mass is clearly too large for a PN.
Therefore, we conclude that the $-$7 km s$^{-1}$ nebula 
photoionized by KPD\,0005+5106 consists of interstellar material.

As a consistency check, we have also evaluated the ionization 
requirement.
The H$\alpha$ luminosity of the ionized gas requires an ionizing
luminosity of $1.4\times10^{46}$ photons s$^{-1}$.
Synthetic stellar fluxes calculated for hot central stars of 
planetary nebulae are available at
http://astro.uni-tuebingen.de/$\sim$rauch/.
Using a helium NLTE model for $T_{\rm eff}$ = 120,000 K and
log $g$ = 7 (in cgs units) and adopting a stellar mass of 0.59 
M$_\odot$ for KPD\,0005+5106 \citep{WHF94}, we find an ionizing
luminosity of $6.5\times10^{45}$ photons s$^{-1}$, within a factor
of 2 from that required to ionize the gas.
The good agreement between the expected and required ionizing
luminosities suggests that the spectroscopic distance of 270 pc 
is reasonably accurate.

\section{Nature of the Hard X-ray Emission from KPD\,0005+5106}

The nature of the spatially-unresolved hard X-ray emission from 
KPD\,0005+5106 is puzzling.
Hard X-ray emission from WDs is usually attributed to their 
binary companions, and a late-type companion can be diagnosed 
by a near-IR excess; however,
KPD\,0005+5106 shows no obvious near-IR excess, with $V$ = 13.32,
$J$ = 13.93, $H$ = 14.13, and $K$ = 14.18 \citep{Oetal03}.
The absence of detectable near-IR excess places an upper limit on
the luminosity and spectral type of a companion.
The hard X-ray luminosity of KPD\,0005+5106 also constrains the
luminosity of a hidden companion, as $L_{\rm X}/L_{\rm bol} \sim
10^{-3} - 10^{-4}$ for coronal X-ray emission from late-type
stars \citep{FSG95}.
Below we use the near-IR and X-ray constraints to assess the
possibility of a hidden late-type companion that is responsible
for the hard X-ray emission from KPD\,0005+5106.

The {\it ROSAT} PSPC observation RF200428 detected a total of 
$25\pm5$ counts (background-subtracted) from KPD\,0005+5106 
in the 0.4--2.0 keV band.
To convert the count rate to luminosity, we use the energy-count 
conversion factors for {\it ROSAT} PSPC observations with a boron 
filter in the hard band ($\ge$0.4 keV) for thermal line spectra
provided in Figure 10.12 of the {\it ROSAT} Mission Description.
We adopt $N_{\rm H} = 5\times10^{20}$ H-atoms cm$^{-2}$ derived 
from Lyman $\alpha$ absorption profile by \citet{WHF94} as the 
absorption column density.
For plasma temperatures of $2-5\times10^6$ K, the conversion
factor is $0.4-3.5\times10^{10}$ counts cm$^2$ erg$^{-1}$.
The unabsorbed hard X-ray flux from KPD\,0005+5106 is thus
$0.14-1.3\times10^{-12}$ ergs cm$^{-2}$ s$^{-1}$.

The full range of uncertainty in the spectroscopic distance to 
KPD\,0005+5106, 270$^{+210}_{-120}$ pc \citep{WHF94}, is 
considered in the discussion below.
Using the near distance (150 pc), the X-ray luminosity of 
KPD\,0005+5106 is $0.4-4\times10^{30}$ ergs s$^{-1}$.
If the hard X-ray emission originates from the corona of a late-type 
companion, its $L_{\rm bol}$ is at least $4\times10^{32} - 
4\times10^{34}$ ergs s$^{-1}$, corresponding to $M_{\rm bol}$ = 
7.2 -- 2.2.
Assuming that the companion is a main sequence star, the least 
luminous possible spectral type is M0.
An M0\,V star at a distance of 150 pc would have $K = 11.1$, and 
would be too luminous for KPD\,0005+5106's $K = 14.18$.
Using the far distance (480 pc), the X-ray luminosity of 
KPD\,0005+5106 is $0.4-4\times10^{31}$ ergs s$^{-1}$.
Using the same derivation, we find the faintest possible spectral
type for a coronal companion is G5.
A G5\,V star at a distance of 480 pc would have $K = 11.7$, and
would also be too luminous to be hidden by KPD\,0005+5106.
Therefore, the near-IR photometric measurements and the hard X-ray 
luminosity of KPD\,0005+5106 exclude the possible existence of a 
late-type companion whose corona contributes to the observed hard 
X-ray emission.
This conclusion is also supported by our echelle observations 
of KPD\,0005+5106, which did not detect any H$\alpha$ emission 
from a late-type companion.

The discovery of a photoionized nebula around KPD\,0005+5106 opens
up another possible mechanism to generate X-ray emission: accretion
of interstellar material onto the stellar surface.
The rate of gravitational energy released by the accreted material
is $\pi G R_* M_* v \rho_0$, where $G$ is the gravitational constant,
$R_*$ is the stellar radius, $M_*$ is the stellar mass, $v$ is the
translational velocity of star through the interstellar gas, and
$\rho_0$ is the density of the ambient interstellar gas.
Adopting $R_* = 2.8\times10^9$ cm and $M_*$ = 0.59 M$_\odot$ as
determined from the spectral analysis of \citet{WHF94}, a 
translational velocity of 40 km s$^{-1}$, and an interstellar 
density of 1 H-atom cm$^{-3}$, we find an accretion luminosity 
$\sim 6\times10^{18}$ ergs s$^{-1}$, more than twelve orders of 
magnitude lower than the hard X-ray luminosity of KPD\,0005+5106.
Therefore, we conclude that the accretion of interstellar material
does not contribute much to the hard X-ray emission.

Is the hard X-ray emission from KPD\,0005+5106 intrinsic or
extrinsic?  
The angular resolution of the {\it ROSAT} PSPC, $\sim$30$''$,
is not able to resolve background sources projected within a 
few arcsec from KPD\,0005+5106.
However, the stellar \ion{O}{8} emission from KPD\,0005+5106 
\citep{Wetal96,Setal97} can be used to argue for a local,
instead of background, origin of the hard X-rays.
The \ion{O}{8} recombination line emission requires the 
ionization of O$^{+7}$, which may be achieved through thermal
collisions in wind outflows or photoionization by X-rays 
at energies greater than 0.871 keV.
The suggestion of a wind outflow of KPD\,0005+5106 was 
based on the velocity offset between the stellar photospheric 
lines and the ``circumstellar'' absorption lines \citep{SD92};
however, we have demonstrated in \S4 that this ``circumstellar''
material is really interstellar medium.
There is therefore no evidence for a wind outflow.
Without a wind outflow, a hard ($>$0.871 keV) X-ray radiation
field is needed, and the source has to be local to KPD\,0005+5106.
We have ruled out the existence of a late-type companion with
a corona; thus, KPD\,0005+5106 must be responsible for the hard
X-ray emission.
Stellar X-ray emission may originate from the photosphere or 
a corona; however, X-ray emission near 1 keV is not expected from
models of the atmosphere of KPD\,0005+5106 and WDs are not 
expected to possess hot coronae.
The hard X-ray emission from KPD\,0005+5106 is most puzzling.

\section{Summary and Conclusions}

We have detected a large ionized nebula encompassing both
KPD\,0005+5106 and AO Cas.
The nebular velocities and [\ion{O}{3}]/H$\alpha$ ratios are
used to separate the high-excitation nebula ionized by 
KPD\,0005+5106 and the background \ion{H}{2} region ionized 
by AO Cas.
The nebula ionized by KPD\,0005+5106 contains interstellar
gas because its radial velocity is similar to that of the local 
interstellar medium and the nebular mass ($\sim$70 M$_\odot$) 
is much higher than those of planetary nebulae.

The interstellar origin of the nebula removes the basis for
the suggestion of a wind outflow from KPD\,0005+5106 -
an offset of about +50 km s$^{-1}$ between the nebular
absorption line velocities and the photospheric line
velocities.
The recently discovered large ionized nebula around the hot DO 
WD PG\,1034+001 \citep{Hetal03} also shows a nebular velocity
similar to the circumstellar/interstellar absorption line velocities 
and the velocity of local interstellar gas but offset by about 
+45 km s$^{-1}$ from the stellar photospheric line velocities 
\citep{Hetal98}.
We suggest that the large ionized nebula around PG\,1034+001
is also interstellar in origin.
The large ionized nebulae of KPD\,0005+5106 and PG\,1034+001 clearly
demonstrate that hot WDs contribute to the ionization of the 
interstellar medium.
The brightest hot WDs are also the nearest, within a few hundred
pc; thus, their photoionized interstellar nebulae would be angularly
large so that the WHAM-NSS provides the best data set to search for
these nebulae.

We have analyzed the hard X-ray luminosity of KPD\,0005+5106.
Using the $L_{\rm X}/L_{\rm bol}$ relation for typical late-type
stars with active coronae and  KPD\,0005+5106's lack of obvious 
near-IR excess, we have excluded the possibility that a late-type
companion's coronal activity is responsible for the hard X-ray
emission from KPD\,0005+5106.
The stellar \ion{O}{8} emission of KPD\,0005+5106 places useful
constraints on the location of the hard X-ray source. 
Our analysis of the nebular environment of KPD\,0005+5106 
abrogates the previous suggestion of a wind outflow from 
KPD\,0005+5106; thus, a local hard X-ray radiation field
is needed to photoionize O$^{+7}$ in order to produce 
the stellar \ion{O}{8} emission.

We conclude that the most likely origin of the hard X-ray 
emission from KPD\,0005+5106 is the DO WD itself, either
from its photosphere or a corona, although such 
photospheric emission is not expected from the currently 
available stellar atmospheric models and there are no known
mechanisms to generate hot coronae around hot WDs.
To further study the origin of hard X-ray emission from 
KPD\,0005+5106, X-ray observations with high angular resolution
and high sensitivity are needed to confirm the positional 
coincidence between the hard X-ray source and KPD\,0005+5106 
and to obtain high-quality data for spectral analysis.
Future observations with {\it Chandra X-ray Observatory} and
{\it XMM-Newton Observatory} are most desirable.

\acknowledgments

This research has used data from the Wisconsin H-Alpha Mapper,
which is funded by the National Science Foundation.
YHC acknowledges the support of NASA grant NAG 5-13076.
X-ray data analysis in T\"ubingen is supported by the DLR under grant
50\,OR\,0201.

\clearpage

\begin{figure}
\figurenum{1}
\epsscale{0.5}
\plotone{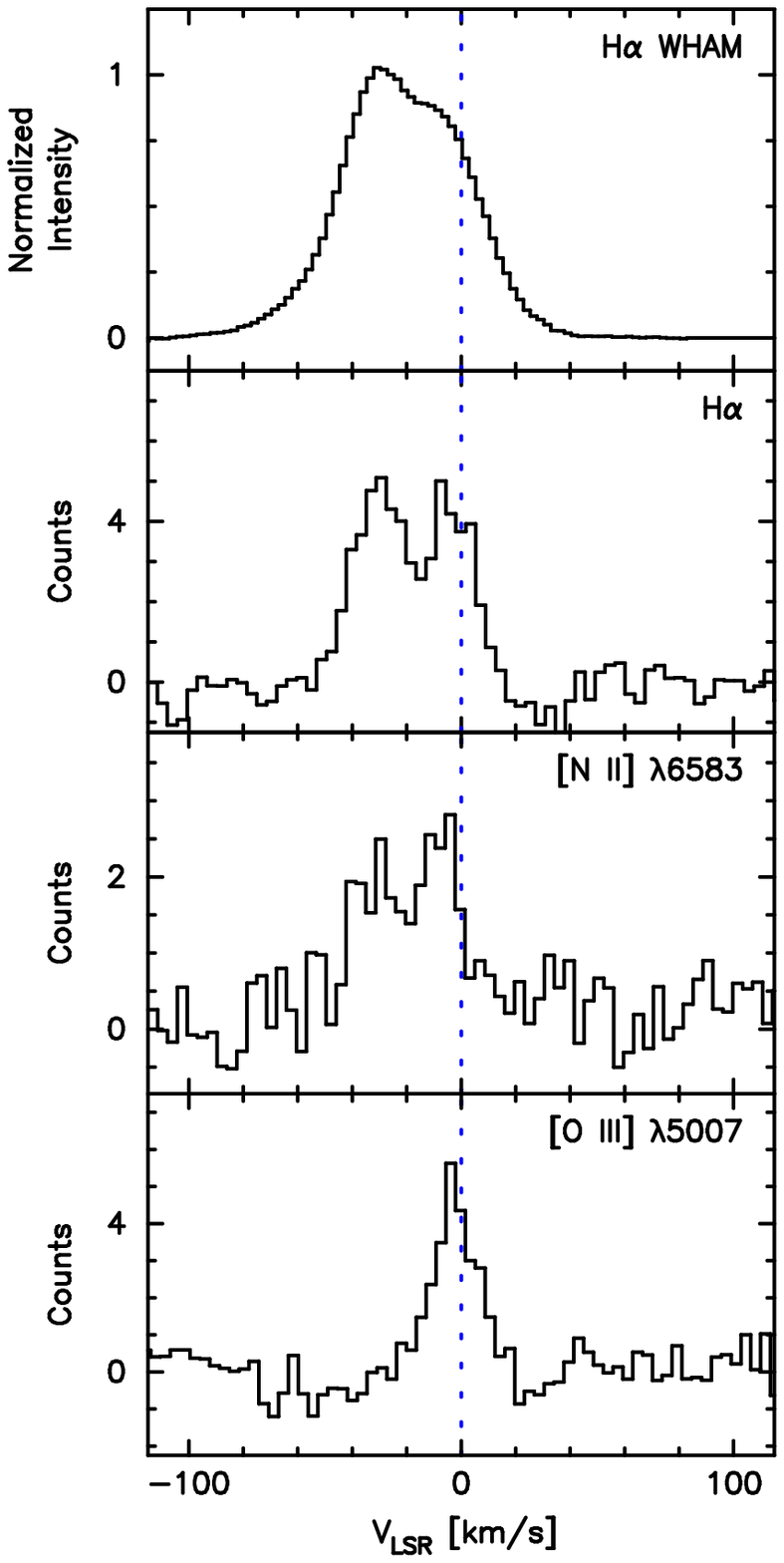}
\caption{Velocity profiles of the H$\alpha$, [\ion{N}{2}],
and [\ion{O}{3}] lines.  The H$\alpha$ line profile in the 
top panel is extracted from the WHAM-NSS and normalized to 
its peak.  The other line profiles are extracted from our 
echelle observations.
}
\label{fig1}
\end{figure}

\begin{figure}
\figurenum{2}
\epsscale{1}
\plotone{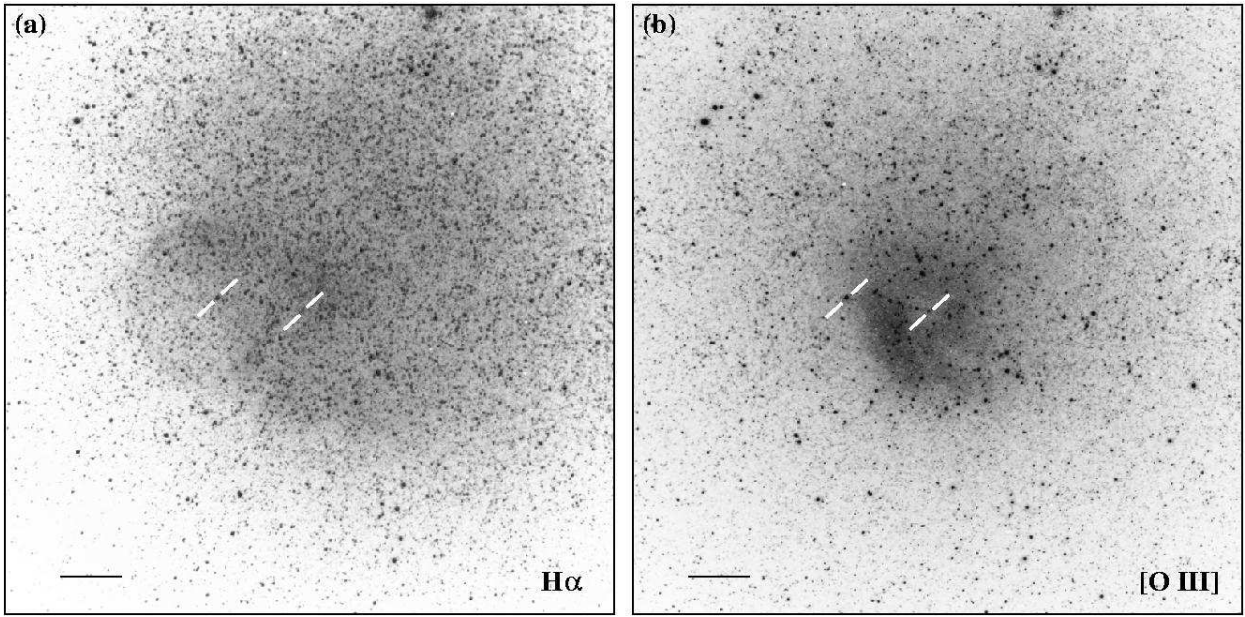}
\caption{(a) H$\alpha$ and (b) [\ion{O}{3}] images from Gull \& Chu
emission-line survey of the northern sky, using a 135-mm-focal-length
lens.  North is up and east is to the left.  The stars AO Cas 
(on the left) and KPD\,0005+5106
(on the right) are marked.  The [\ion{O}{3}]-emission region is 
roughly centered at KPD\,0005+5106. The 1$^\circ$ scale is marked 
by the horizontal line at the lower left corner of each panel.
}
\label{fig2}
\end{figure}

\begin{figure}
\figurenum{3}
\epsscale{1}
\plotone{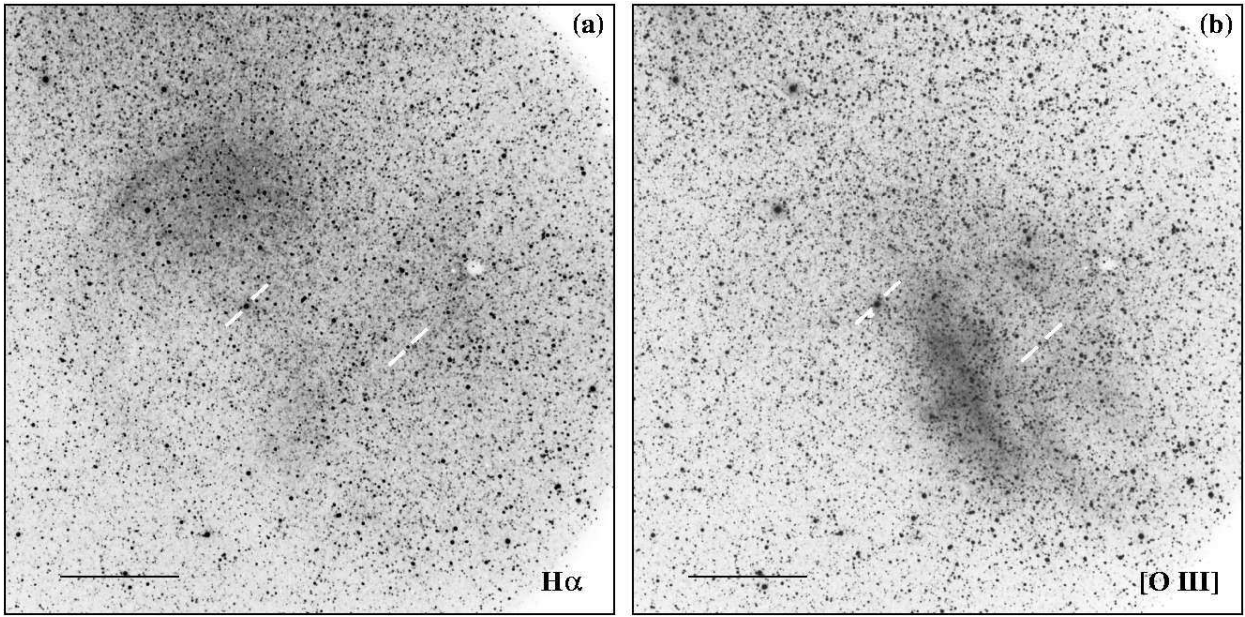}
\caption{(a) H$\alpha$ and (b) [\ion{O}{3}] images from follow-up
observations of An Emission-Line Survey of the Milky Way
\citep{PGK79},  using a 300-mm-focal-length lens. 
The [\ion{O}{3}]-emission region is 
roughly centered at KPD\,0005+5106. The 1$^\circ$ scale is marked 
by the horizontal line at the lower left corner of each panel.
}
\label{fig3}
\end{figure}

\begin{figure}
\figurenum{4}
\plotone{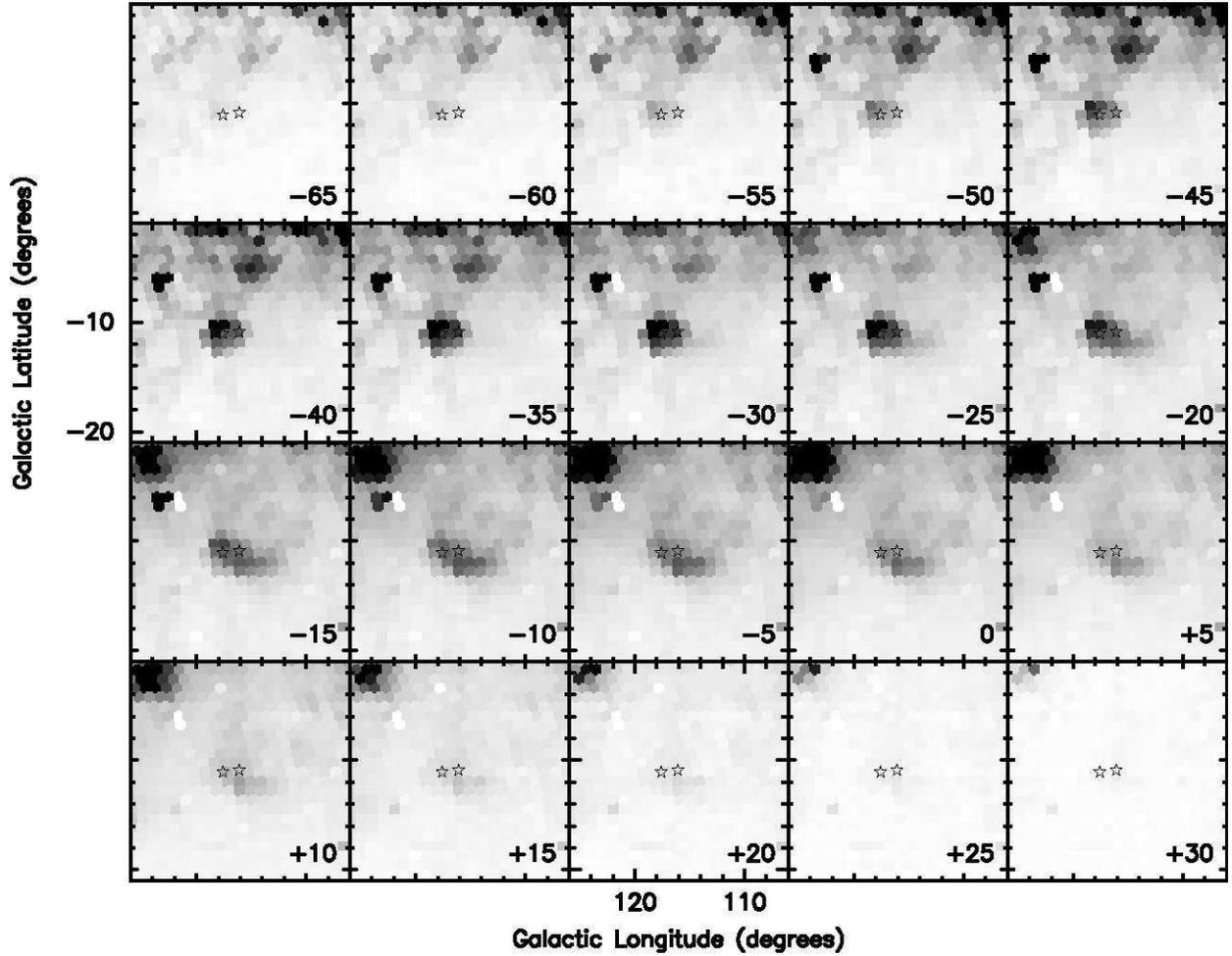}
\caption{Iso-velocity maps extracted from the WHAM-NSS data.
The local standard of rest velocity ($V_{\rm LSR}$) of each
panel is marked at the lower right corner.  The two stars 
marked are AO Cas (on the left) and KPD\,0005+5106 (on the right).
These maps are presented in the galactic coordinates,
so the orientation is rotated counterclockwise by about 15$^\circ$
with respect to that of Figure 2.
}
\label{fig4}
\end{figure}


\begin{thebibliography}{}

%\bibitem[Bohlin et al.(1978)Bohlin, Savage, \& Drake]{BSD78}
% Bohlin, R.~C., Savage, B.~D., \& Drake, J.~F.\ 1978, \apj, 224, 132 

\bibitem[Fleming et al.(1995)Fleming, Schmitt, \& Giampapa]{FSG95} 
 Fleming, T.~A., Schmitt, J.~H.~M.~M., \& Giampapa, M.~S.\ 1995, 
 \apj, 450, 401 

\bibitem[Fleming et al.(1993)Fleming, Werner, \& Barstow]
{FWB93} Fleming, T.\ A., Werner, K., \& Barstow, M.\ A.\ 1993,
\apjl, 416, L79

\bibitem[Gies \& Wiggs(1991)]{GW91} Gies, D.~R.~\& Wiggs, 
M.~S.\ 1991, \apj, 375, 321 

\bibitem[Gruendl et al.(2001)]{Getal01} Gruendl, R.~A., Chu, Y.-H.,
 O'Dwyer, I.~J., \& Guerrero, M.~A.\ 2001, \aj, 122, 308 

\bibitem[Haffner et al.(2003)]{H03} 
 Haffner, L.~M. et al.\ 2003, \apjs, 149, 405

\bibitem[Hartmann \& Burton(1997)]{HB97} Hartmann, D.,~\& Burton,
 W.~B.\ 1997, Atlas of Galactic Neutral Hydrogen, Cambridge 
University Press, ISBN 0521471117

\bibitem[Hewett et al.(2003)]{Hetal03} Hewett, P.~C., et al.\
 2003, \apjl, 599, L37 

\bibitem[Holberg et al.(1998)Holberg, Barstow, \& Sion]{Hetal98} 
Holberg, J.~B., Barstow, M.~A., \& Sion, E.~M.\ 1998, \apjs, 119, 207 

\bibitem[Kruk \& Werner(1996)]{KW96} Kruk, J.\ W., \& Werner,
K.\ 1996, in H-Deficient Stars, eds.\ U.\ Heber \&  C.\ S.\ Jeffery,
the ASP Conference Series, 96, 205

\bibitem[Kruk \& Werner(1998)]{KW98} Kruk, J.\ W., \& Werner,
 K.\ 1998, \apj, 502, 858

%\bibitem[Moos et al.(2000)]{moos00} 
%Moos, H.~W.~et al.\ 2000, \apjl, 538, 1

%\bibitem[Nicolet(1978)]{Ni78} Nicolet, B.\ 1978, \aaps, 34, 1

\bibitem[O'Dwyer et al.(2003)]{Oetal03} O'Dwyer, I.\ J., Chu,
 Y.-H., Gruendl, R.\ A., Guerrero, M.\ A., \& Webbink, R.\ F.\
 2003, \aj, 125, 2239

\bibitem[Otte et al.(2004)Otte, Dixon, \& Sankrit]{ODS04}
Otte, B., Dixon, W.\ V. D., \& Sankrit, R.\ 2004, \apjl, 606, L143

\bibitem[Parker et al.(1979)Parker, Gull, \& Kirshner]{PGK79}
 Parker, R.\ A.\ R., Gull, T.\ R., \& Kirshner, R.\ P.\ 1979,
 An Emission-Line Survey of the Milky Way, NASA SP-43A

\bibitem[Rauch et al.(2004)Rauch, Kerber, \& Pauli]{Retal04} Rauch, 
T., Kerber, F., \& Pauli, E.-M.\ 2004, \aap, 417, 647 

\bibitem[Reynolds(1987)]{R87} Reynolds, R.~J.\ 1987, \apj, 315, 234 

\bibitem[Reynolds(1988)]{R88} Reynolds, R.~J.\ 1988, \apj, 333, 341 

%\bibitem[Sahnow et al.(2000)]{sahnow00}
%Sahnow, D.~J.~et al.\ 2000, \apjl, 538, 7 

\bibitem[Sion \& Downes(1992)]{SD92} Sion, E.~M.~\& Downes, 
R.~A.\ 1992, \apjl, 396, L79 

\bibitem[Sion et al.(1997)]{Setal97} Sion, E.\ M., Holberg, J.\ 
 B., Barstow, M.\ A., \& Scheible, M.\ P.\ 1997, \aj, 113, 364

\bibitem[Walborn(1973)]{W73} Walborn, N.~R.\ 1973, \aj, 78, 1067

\bibitem[Werner et al.(1997)]{Wetal97} Werner, K., Bagschik, K.,
 Rauch, T., \& Napiwotzki, R.\ 1997, \aap, 327, 721

\bibitem[Werner et al.(1994)Werner, Heber, \& Fleming]{WHF94}
 Werner, K., Heber, U., \& Fleming, T.\ 1994, \aap, 284, 907

\bibitem[Werner et al.(1996)]{Wetal96} Werner, K., et al.\
1996, \aap, 307, 860

\bibitem[Wilson(1953)]{W53} Wilson, R.~E.\ 1953, 
General Catalogue of Stellar Radial Velocities, Carnegie 
Institute Washington D.C.~Publication, 601

\end{thebibliography}
\end{document}